\documentstyle[sprocl,psfig]{article}

\def\Journal#1#2#3#4{{#1} {\bf #2}, #3 (#4)}

\def\be{\begin{equation}}
\def\ee{\end{equation}}
\def\bea{\begin{eqnarray}}
\def\eea{\end{eqnarray}}

\def\gtsima{$\, \buildrel > \over \sim \,$}
\def\simgt{\lower.5ex\hbox{\gtsima}}
\def\ltsima{$\, \buildrel < \over \sim \,$}
\def\simlt{\lower.5ex\hbox{\ltsima}}

\begin{document}

\title{Keck Spectroscopy of Dwarf Elliptical Galaxies in the Virgo Cluster}

\author{M. Geha, P. Guhathakurta}

\address{UCO/Lick Observatory, Santa Cruz, CA 95064, USA\\E-mail: mgeha, raja@ucolick.org} 

\author{R. van der Marel}

\address{STScI, 3700 San Martin Drive, Baltimore, MD 21218, USA\\E-mail: marel@stsci.edu}


\maketitle\abstracts{Keck spectroscopy is presented for four dwarf
elliptical galaxies in the Virgo Cluster.  At this distance, the mean
velocity and velocity dispersion are well resolved as a function of radius
between 100 to 1000~pc, allowing a clear separation between
nuclear and surrounding galaxy light.  We find a variety of dispersion
profiles for the inner regions of these objects, and show that none of
these galaxies is rotationally flattened.}

\section{Introduction}
Dwarf elliptical galaxies (dEs) are among the poorest studied galaxies
due to their faint luminosities, $M_V \simgt -17$, and characteristic
low effective surface brightness $\mu_e(V) > 22$~mag~arcsec$^{-2}$
(Ferguson \& Binggeli 1994).  The defining characteristic of dEs is an
exponential surface brightness profile.  The majority of dEs brighter
than $M_V= -16$ have compact nuclei typically containing 5 to 20\%
of the total galaxy light; most dEs fainter than $M_V = -12$ show no
sign of a nucleus (Sandage {\it et al.} 1985).  Although the sample of
dEs with measured internal kinematics is small (Bender \&
Nieto 1990; Peterson \& Caldwell 1993), these observations have
provided strong evidence that dwarf and classical ellipticals evolve
via very different physical processes.

Here we present Keck spectroscopy for four Virgo dEs.  Velocity and
velocity dispersion profiles are measured out to $\sim1$~kpc, assuming
a Virgo Cluster distance of 16.1~Mpc (Kelson {\it et al.}  2000).  These are
the initial results of a larger project to study the dynamics of dwarf
elliptical galaxies.

\section{Keck Observations}
Four Virgo dEs were observed with the Echelle Spectrograph and Imager (ESI)
on the Keck~II telescope in March 2001.  The spectra were obtained through a
$0.75'' \times 20''$ slit placed along the major axis of each galaxy
with wavelength coverage $\rm\lambda\lambda3900-9500\AA$ and resolution of
23~km~s$^{-1}$ (Gaussian sigma).  As shown in Table~1, the observed galaxies
cover a range of ellipticities and three of the four are nucleated dwarfs
(dE,N).  These objects lie near the bright end of the dE luminosity
function and were selected to have archival WFPC2 imaging.  Mean radial
velocities and velocity dispersions were determined using a pixel space
$\chi^2$ minimization scheme described in van der Marel (1994).
The data were spatially rebinned to achieve a $\rm S/N > 5$ at all radii.
Velocities are measured relative to a K0III template star using the Mg\,b
region, $\rm\lambda\lambda5000-5400\AA$; an analysis of the full wavelength
region will be presented in a forthcoming paper.  Tests show that the
galaxies' internal velocity dispersions are recovered accurately down to the
instrumental resolution of 23~km~s$^{-1}$.

\begin{figure}
\hskip 0.5in
\begin{minipage}{2 in}
\psfig{figure=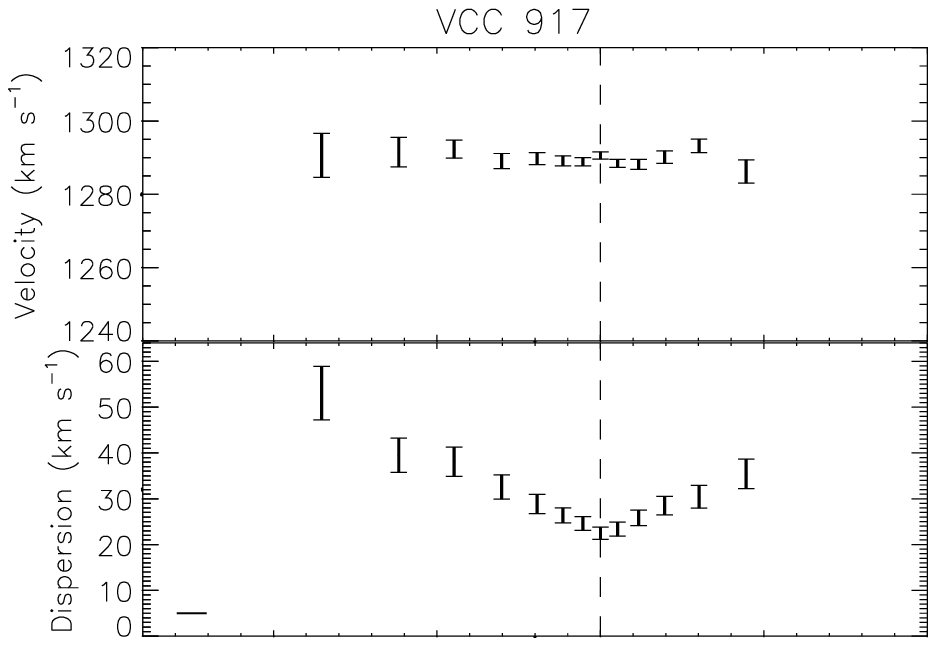,height=1.35in,angle=90}
\end{minipage}
\hskip 0.4in
\begin{minipage}{2 in}
\psfig{figure=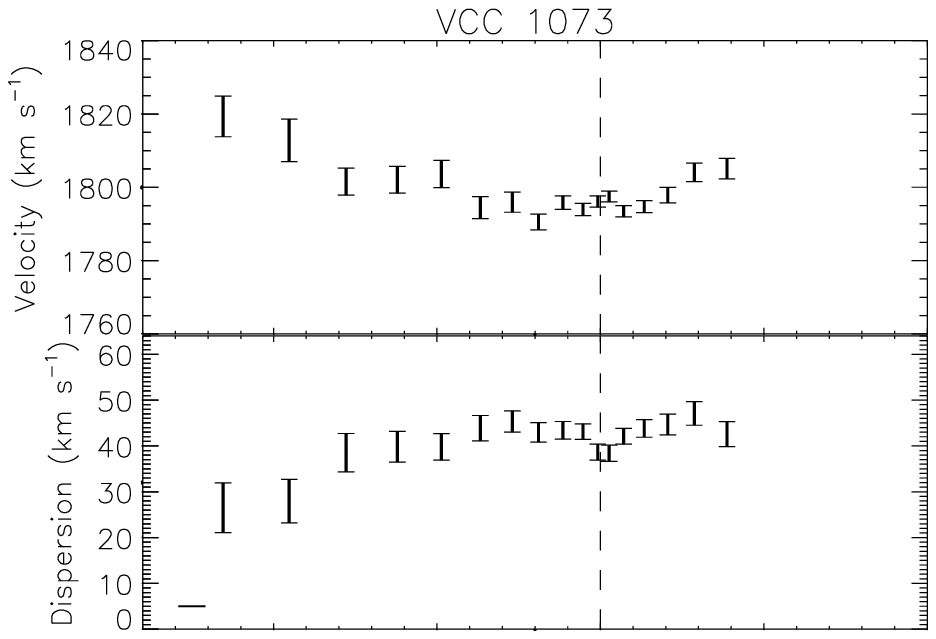,height=1.35in,angle=90}
\end{minipage}

\vskip 0.2 in
\hskip 0.5in
\begin{minipage}{2 in}
\psfig{figure=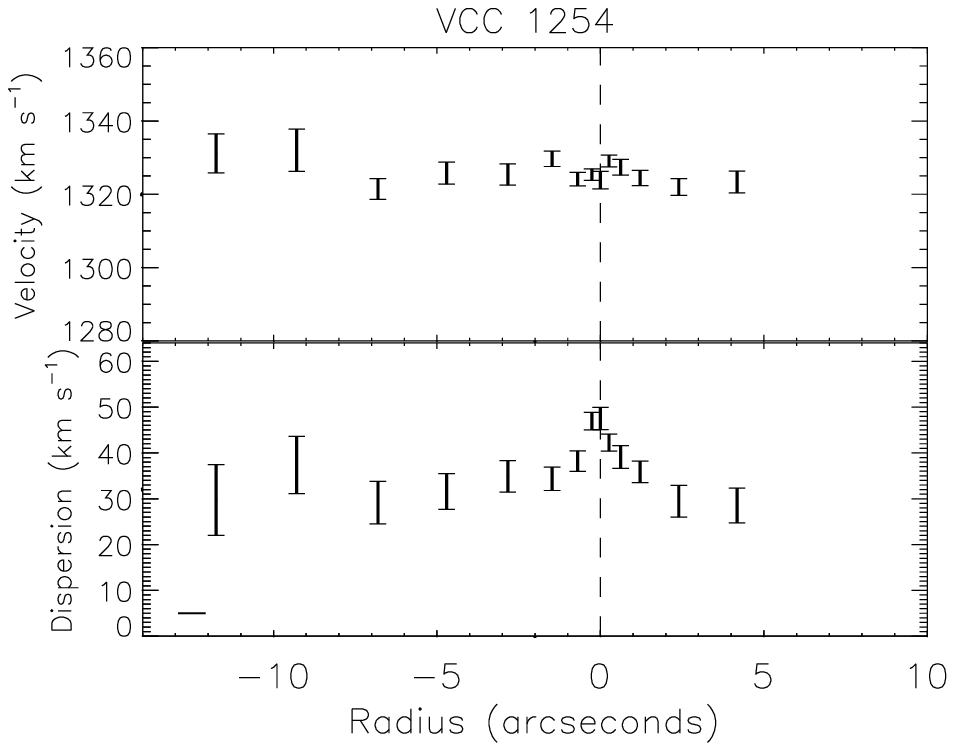,height=1.35in,angle=90}
\end{minipage}
\hskip 0.4in
\begin{minipage}{2 in}
\psfig{figure=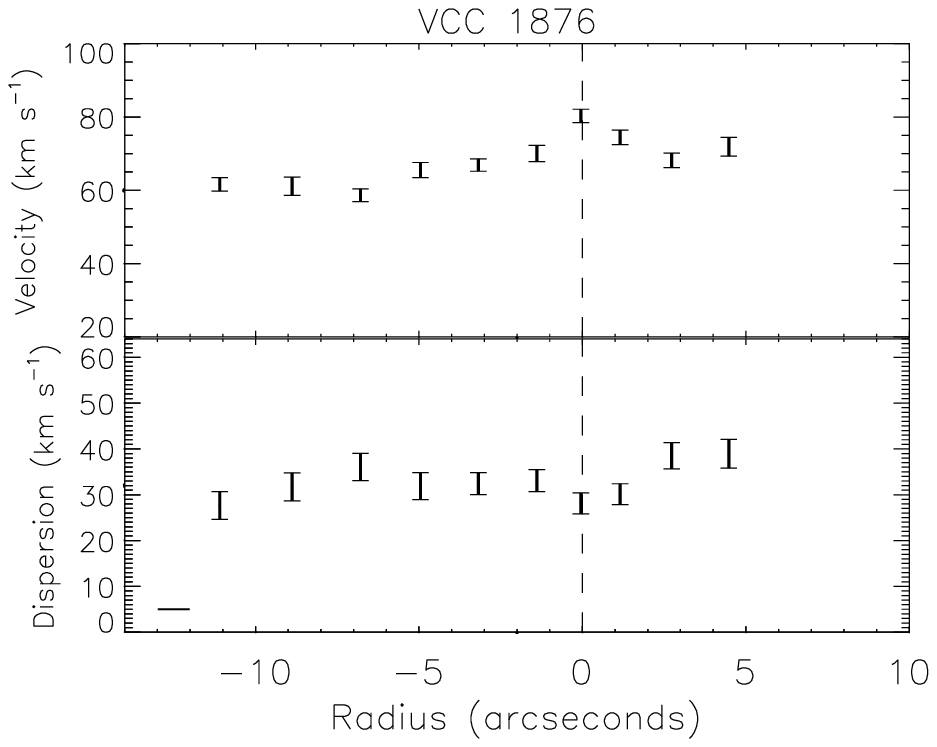,height=1.35in,angle=90}
\end{minipage}

\vskip 0.2in
\caption{Mean velocity and velocity dispersion profiles for four Virgo
dEs. The bar at the lower left of each panel indicates the seeing FWHM
during each observation.  At the distance of the Virgo Cluster,
$1'' \sim 100$~pc. }
\end{figure}

\section{Discussion}
\subsection{Anisotropic Dispersion Versus Rotational Flattening}
The observed shapes and kinematics of elliptical galaxies between $-20
< M_B < -18$ are consistent with rotational flattening.  This trend
does not appear to extend to lower luminosity classical ellipticals
and the three Local Group dwarf ellipticals (Davies {\it et al.} 1983;
Bender \& Nieto 1990).  The four Virgo dEs presented here are also not
rotationally flattened.  For each galaxy, an average ellipticity
$\epsilon$ was determined by standard ellipse fitting of archival
WFPC2 $V$-band images between radii of $1''-20''$ (see Table 1).  From the
velocity profiles shown in Figure~1, we estimate an upper limit to the
maximum rotation velocity, $v_{\rm max}$.  An average velocity dispersion
$\sigma$ is determined for each galaxy beyond $r>1''$ to avoid nuclear
contamination.

The ratio $v_{max}/\sigma$ is plotted against ellipticity in Figure~2
and is compared to the ratio expected from an isotropic, rotationally
flattened body (Binney \& Tremaine 1987).  The upper limits on
$v_{max}/\sigma$ determined for these galaxies are 2 to 8 times smaller
than expected if the observed flattenings were due to rotation.  Thus, we
conclude that these dEs are primarily flattened by anisotropic velocity
dispersions.

\begin{figure}[t]
\hskip 0.5 in
\psfig{figure=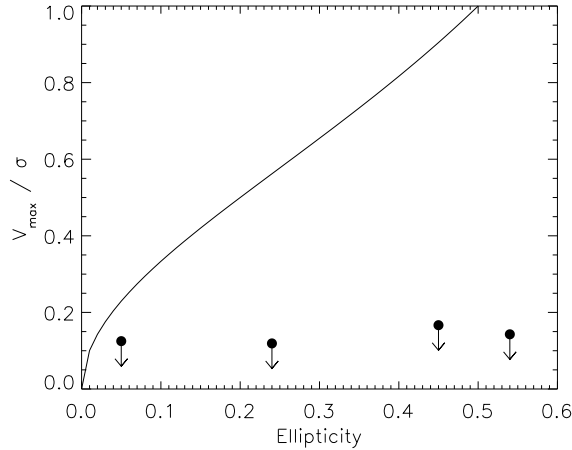,height=2.6in,angle=90}
\caption{The ratio of the upper limit on the rotation velocity $v_{\rm max}$
to observed velocity dispersion $\sigma$ plotted versus mean
ellipticity for four Virgo dwarf ellipticals.  The solid line is 
the expected relation for an oblate, isotropic galaxy flattened
by rotation.}
\end{figure}

\subsection{Velocity Dispersion Profiles and dE Nuclei}
Although the mean velocity profiles presented in Figure~1 are
qualitatively similar, the velocity dispersion profiles are more
heterogeneous.  The velocity dispersion of the non-nucleated dE VCC~917
decreases smoothly towards the galaxy center in contrast to the
three nucleated dwarfs, which vary more abruptly in the central few
arcseconds.  The nuclear velocity dispersions of two dE,Ns, VCC~1073 and
VCC~1876, are lower than the surrounding galaxy, whereas the nuclear
velocity dispersion of VCC~1254 is higher.  The origin of nuclei in dEs is
largely unknown, but their presence has been correlated to global
galaxy parameters such as shape and specific globular cluster
frequency (Ryden \& Terndrup 1994; Miller {\it et al.} 1998).  A
favored hypothesis is that the nuclei are dense star clusters,
possibly remnants of larger stripped or harassed objects (Moore {\it
et al.} 1998).  More work is needed to determine whether the kinematic
profiles presented here are consistent with these scenarios.

Dynamical mass modeling of classical ellipticals has placed strong
constraints on their origin and evolution.  We are in the process of
modeling these dE kinematic data using techniques similar to those
described in van der Marel (1994).  We will investigate the variation of
$M/L$ ratio across our sample of galaxies and as a function of galactic
radius within each galaxy.
In addition, we plan to study the position of these dEs in the
Fundamental Plane.  These results will be presented in an forthcoming
paper.

\begin{table}[t]
\caption{Observed Virgo Cluster Dwarf Elliptical Galaxies}
\vspace{0.2cm}
\begin{center}
\footnotesize
\begin{tabular}{l c c c} \hline
\multicolumn{1}{c}{Galaxy Name}&
\multicolumn{1}{c}{$M_B$}&
\multicolumn{1}{c}{Type}&
\multicolumn{1}{c}{$\epsilon$}\\ \hline
VCC 917  & $-16.4$ & dE6   & 0.54\\
VCC 1073 & $-17.3$ & dE3,N & 0.24\\
VCC 1254 & $-16.4$ & dE0,N & 0.05\\
VCC 1876 & $-16.8$ & dE5,N & 0.45\\
\hline
\end{tabular}
\end{center}
\end{table}

\end{document}